\documentstyle[epsf,rotate]{elsart}

\begin{document}

\begin{frontmatter}
\title{The first 20 minutes in the Hong Kong stock market}
\author{Zhi-Feng Huang\thanksref{email}}
\address{Institute for Theoretical Physics, Cologne University,
D-50923, K\"oln, Germany}
\thanks[email]{{\em E-mail address}: zfh@thp.uni-koeln.de}
\maketitle

\begin{abstract}
Based on the minute-by-minute data of the Hang Seng Index in
Hong Kong and the analysis of probability distribution and 
autocorrelations, we find that the index fluctuations for the 
first few minutes of daily opening show behaviors very different
from those of the other times. In particular, the properties 
of tail distribution, which will show the power law scaling
with exponent about $-4$ or an exponential-type decay, the 
volatility, and its correlations depend on the opening effect 
of each trading day.
\end{abstract}
\begin{keyword}
Probability distribution; Volatility; Autocorrelation;
Exponential; Power law.
\end{keyword}

\end{frontmatter}

\section{Introduction}

Recently, detailed analysis on the high-frequency financial
market data has shown that there exist some universal statistical
characteristics for price or index fluctuations, in particular
the fat tail distribution and rapid decay of correlation for 
price changes, and the persistence of long-range volatility
correlation \cite{bp-book,ms-book,lux1,gpams}. For one of
these fundamental features, the probability distribution,
the power-law asymptotic behavior with an exponent about $-4$
has been found from the daily and high-frequency intra-daily
stock market data \cite{lux1,gpams}.

Many efforts have been made to simulate the market behaviors 
and dynamics, and then to reproduce these stylized observations 
of real markets. Much work focuses on the microscopic discrete 
models \cite{lls,c-b,lux-m,s-s,huang2}, with different mechanisms 
based on the intrinsic structure of financial markets, including 
the herding and imitation behaviors \cite{c-b,s-s} as well as
the mutual interactions \cite{lux-m} among market participants.

The other way to model the dynamics of financial markets is
using the approach of continuous stochastic process and then,
e.g., determining the effective stochastic equation for price
evolution \cite{t-h,fpr,v-s}. Based on the analysis of Hang Seng 
Index (HSI) in Hong Kong and the method of conditional averages
proposed for generic stationary random time series and previously
applied in fluid turbulence \cite{ching}, a Langevin equation 
reproducing well both the observed probability distribution
of index moves with fat tails and the fast decay of moves
correlation has been derived \cite{t-h}. The existence of a viscous 
market restoring force and a move-enhanced noise is shown in the 
equation. Moreover, an analytic form for the whole range of 
probability distribution has been obtained, and interestingly, the 
corresponding asymptotic tail behavior is an exponential-type decay:
\begin{equation}
P(x)\sim \exp(-\alpha |x|)/|x|,
\label{exp}
\end{equation}
where the index move $x(t)={\rm index}(t)-{\rm index}(t-\Delta t)$ 
with time interval $\Delta t$ (e.g., 1 min), faster than the
power law behavior with exponent about $-4$ found in recent
studies \cite{lux1,gpams,lux-m,s-s}. The parameters can be 
directly determined from the market data (in which the first
20 minutes in the opening of each day are skipped), and the 
tail behavior (\ref{exp}) has also been observed in the simulations 
of our self-organized microscopic model \cite{huang2} with social 
percolation process \cite{solomon,huang1}, which is proposed to 
describe the information spread for different trading ways across 
a social system.

Instead of describing the details of our modelings for financial
market behaviors which have been or will be published elsewhere
\cite{huang2,t-h}, here we present our work on the analysis of 
Hang Seng Index (HSI), showing that the properties of probability 
distribution and volatility correlations for index fluctuations 
depend on the opening effect of each trading day (i.e., the 
overnight effect), which can also explain the above difference 
between the exponential-type fat tail behavior derived in our 
Langevin approach \cite{t-h} and the recent empirical findings 
of $-4$ power law distribution \cite{lux1,gpams}.

\section{Probability distribution}

\begin{figure}
\centerline{\epsfxsize=10cm{\rotate[r]{\epsfbox{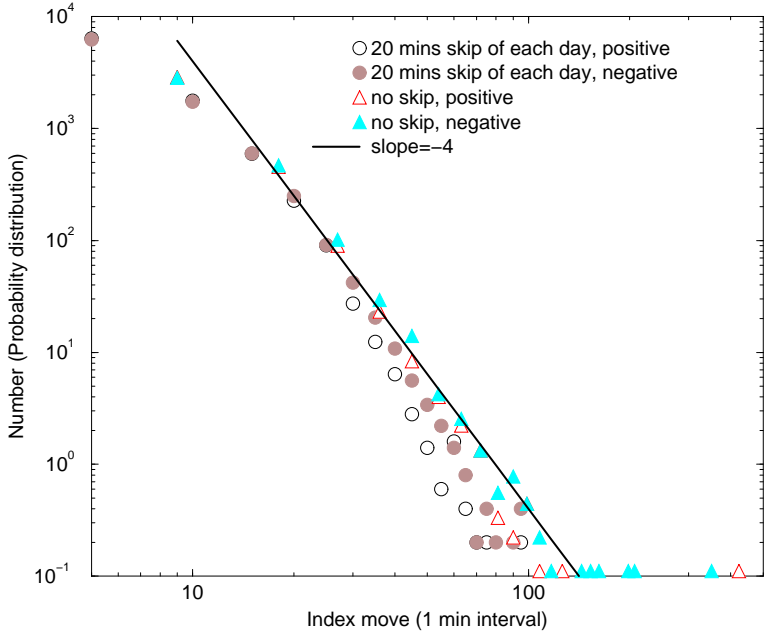}}}}
\caption{Log-log plot of the probability distribution of 1 min
index moves for the Hang Seng Index (HSI) from 1994 to 1997
(open: positive tails, filled: negative tails). The distributions 
with the skip of first 20 minutes in daily opening (circles) and 
without skip (triangles) are shown.}
\label{fig-pdf}
\end{figure}

The HSI data we used contains minute-by-minute records of every
trading day from January 1994 to December 1997, and the break
between the morning and afternoon sessions as well as the
difference between trading days are considered in our analysis.
First, we skip the data in the first 20 minutes of each morning 
session, i.e., skip the opening of each trading day, and the 
deviation from $-4$ power law in the tail region of the distribution 
for 1 min interval index moves is found (Fig. \ref{fig-pdf}, 
circles). In this case the $-4$ power law seems to be a crossover 
effect within a limited range, and for large index moves
the log-log plot exhibits curvature, corresponding to the 
exponential-type Eq. (\ref{exp}) derived from the Langevin 
approach \cite{t-h}.

Next, we analyze the data without any skip in daily opening, 
and it is interesting to find that the $-4$ power law scaling
is recovered for 1 min interval, as shown in Fig. \ref{fig-pdf} 
(triangles), which is in agreement with recent observations from 
German share price index DAX \cite{lux1} and S\&P500 index 
\cite{gpams}.

This phenomenon shows the importance of the daily opening or 
overnight effect for the properties of stock market. It is
well known that price fluctuations in the opening of trading
day are highly influenced by exogenous factors, and the
studies on trading volume have exhibited the larger and less 
elastic transactions demand at opening and close times 
compared with that at other times of the trading day \cite{b-k92}.
Very recently, it has been observed from the German DAX data
that due to the peculiarity in the calculation of the opening index
with the mixture of overnight and high-frequency price changes,
the first observations of each day are governed by the process
different from that of the other times \cite{lux2}. However,
a power law scaling with exponent between 4 and 5 is found
for DAX data in \cite{lux2} when the first 15 minutes of every 
day are dropped, instead of the exponential-type behavior 
here.

\begin{figure}
\centerline{\epsfxsize=10cm{\rotate[r]{\epsfbox{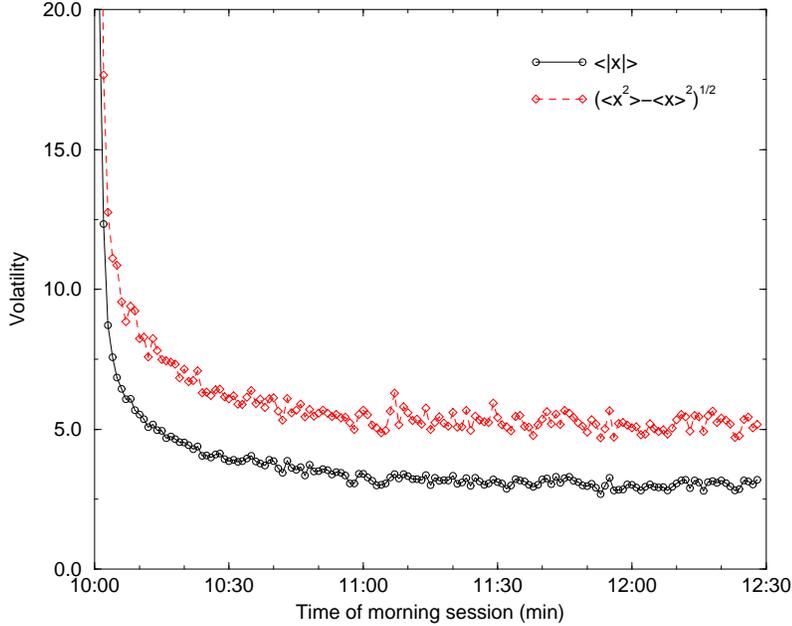}}}}
\caption{The mean of the absolute value of index moves $x$ and the
volatility for different times of morning session (open at 10:00)
in Hong Kong stock market, where the averages are over different 
trading days at the same time.}
\label{fig-volat}
\end{figure}

\section{Volatility and autocorrelations}

For HSI data it is found that the values of index moves and the
volatility at the daily opening times are much larger than 
those of other times. Fig. \ref{fig-volat} shows the mean of the 
absolute value of index moves $\langle |x| \rangle$ and the
volatility $(\langle x^2 \rangle -\langle x \rangle^2)^{1/2}$
for different times of morning session (open at 10:00), where 
the averages are over different trading days from 1994 to 1997 
at the same minute. 
Both of the values are obviously larger for the first 20 minutes,
and then remains almost unchanged at late times, similar to the 
phenomena of German DAX data \cite{lux2}. Thus, when skipping the 
opening data, much less extreme values of index move are calculated
in the probability distribution, and consequently, the far tail 
of distribution may decay faster, as seen in Fig. \ref{fig-pdf}.

\begin{figure}
\centerline{\epsfxsize=10cm{\rotate[r]{\epsfbox{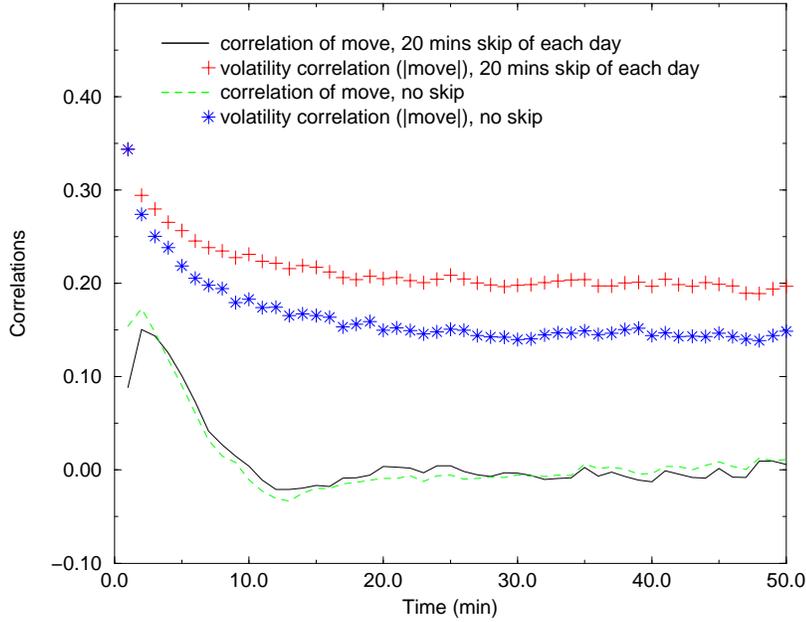}}}}
\caption{Autocorrelations of the index moves 
and the absolute value of index moves (volatility correlations) for
HSI data.}
\label{fig-corr}
\end{figure}

\begin{figure}
\centerline{\epsfxsize=10cm{\rotate[r]{\epsfbox{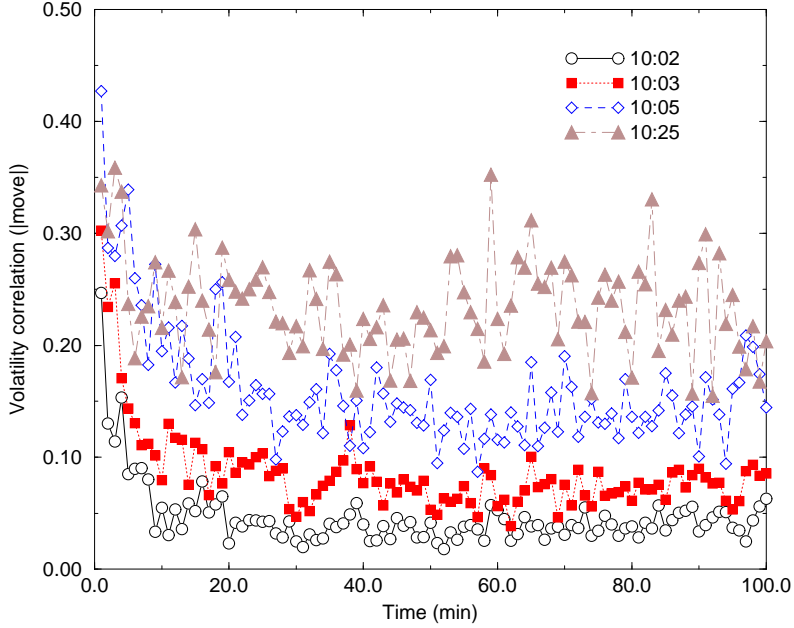}}}}
\caption{Volatility correlations (for the absolute value of index moves) 
for different times: 10:02, 10:03, 10:05, and 10:25 of Hong Kong stock
market.}
\label{fig-open}
\end{figure}

Here we find that the different behavior of distribution shown 
in Fig. \ref{fig-pdf} is relevant to the different properties
of volatility clustering. Fig. \ref{fig-corr} shows the 
autocorrelations of index moves and volatility for 1 min 
interval, with and without the skip of first 20 minutes, where 
the correlation for the index move $x$ 
\begin{equation}
C(T)=\frac {\langle x(t)x(t+T) \rangle-\langle x(t)\rangle^2}
{\langle x(t)^2\rangle-\langle x(t)\rangle^2}
\label{linear}
\end{equation}
rapidly decays to zero in about 10 minutes, and the persistence
of long-range volatility correlation,
\begin{equation}
V(T)=\frac {\langle |x(t)||x(t+T)| \rangle-\langle |x(t)|\rangle^2}
{\langle |x(t)|^2\rangle-\langle |x(t)|\rangle^2},
\label{volat}
\end{equation}
(averaged over the whole index time series)
is found, in accordance with the previous studies 
\cite{bp-book,ms-book}. The correlations of moves
present little difference with or without the skip, however,
the volatility correlation with no skip (Fig. \ref{fig-corr}, 
stars) is obviously smaller. This decrease is due to the fact 
that the volatility correlations of the first few minutes in the
daily opening are much smaller than those of other times, as given 
in Fig. \ref{fig-open}. Note that Hong Kong stock market opens
at 10:00 in the morning, and Fig. \ref{fig-open} shows the volatility 
correlations of different times, defined as 
\begin{equation}
V(t_o,T)=\frac {\langle |x(t_o)||x(t_o+T)| \rangle-
\langle |x(t_o)|\rangle \langle |x(t_o+T)|\rangle}
{\langle |x(t_o)|^2\rangle-\langle |x(t_o)|\rangle^2},
\label{volat-open}
\end{equation}
which is similar to Eq. (\ref{volat}), but averaged only over 
different days (at the same time $t_o$) in the period of 
1994-1997. In the opening time region, 
the value of correlation increases with the increasing of
time, and after the opening (about 20 minutes, i.e., 10:20), 
the correlation keeps relatively unchanged (with the values 
around the pluses of Fig. \ref{fig-corr}).

\begin{figure}
\centerline{\epsfxsize=9.5cm{\rotate[r]{\epsfbox{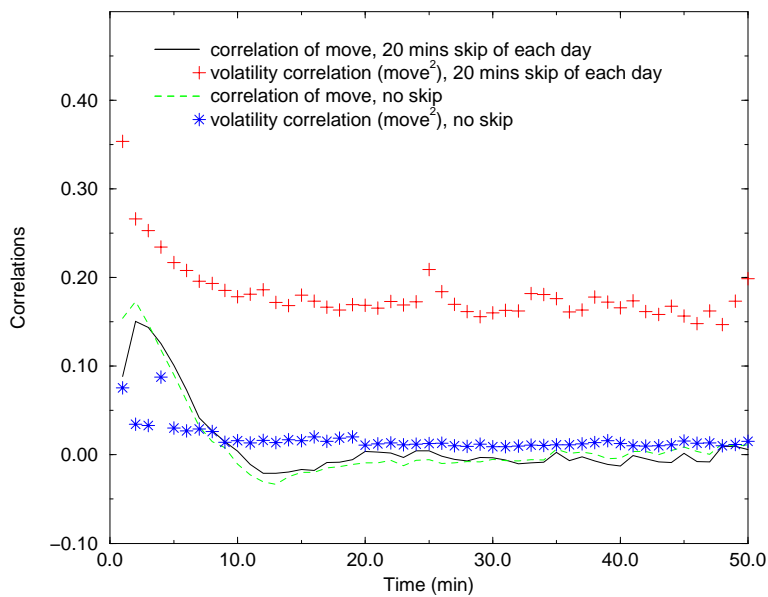}}}}
\caption{Autocorrelations of the square of index moves 
(volatility correlations) for HSI data. Correlations of
index moves are also shown for comparison.}
\label{fig-corr2}
\end{figure}

\begin{figure}
\centerline{\epsfxsize=9.5cm{\rotate[r]{\epsfbox{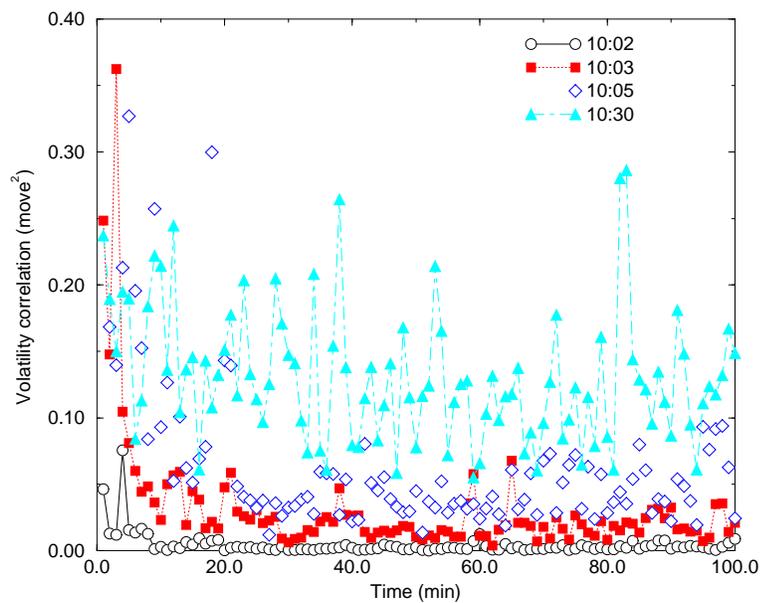}}}}
\caption{Volatility correlations (for the square of index moves) 
for different times: 10:02, 10:03, 10:05, and 10:30 of Hong Kong stock
market.}
\label{fig-open2}
\end{figure}

The absolute value of index move is used to calculated the
volatility correlations in the above study, as shown in
Eqs. (\ref{volat}) and (\ref{volat-open}). If using the
square of move instead, the values of correlation are found 
to be smaller, but the above results will not change, as shown
in Figs. \ref{fig-corr2} and \ref{fig-open2}.

It is known that the Hong Kong stock market behaved abnormally
during the second half of 1997, due to the much more significant
impact of external conditions. When we discard the data of 1997 
and only study the market from 1994 to 1996, the results
are the same as above.

\section{Summary}

In this work we have presented that the index fluctuations for 
the first few minutes of
daily opening behave very differently from those of the other
times, and the lower degree of volatility clustering at the
opening can affect the behaviors of fat tail distribution: 
$-4$ power law behavior if including the daily opening data, 
or the exponential-type if not. To further understand these 
properties of HSI market data, more work is needed to study 
the details of the opening procedure of stock market.

\section*{Acknowledgements}

The author thanks the workshop organizers of "Economic Dynamics 
from the Physics Point of View" for such a very enjoyable
seminar, and Dietrich Stauffer, Lei-Han Tang, and Thomas Lux
for very helpful discussions and comments. I also thank Lam
Kin and Lei-Han Tang for providing the HSI data. This work was 
supported by SFB 341.

\end{document}